\documentclass[twocolumn,showpacs,showkeys,preprintnumbers,amsmath,amssymb]{revtex4}

\usepackage{amsmath}
\usepackage{amssymb}
\usepackage{amsthm}
\usepackage{graphicx}
\usepackage{upref}

\newtheorem{thm}{Theorem}[section]
\newtheorem{cor}{Corollary}[section]

\newtheorem{rk}{Remark}[section]
\newcommand\Cal[1]{{\cal #1}}

\newcommand\complex{\mathbb C}

\newcommand\w{{\rm quant}}

\newcommand\comp{{\rm comp}}

\def\a{{\alpha}}
\def\e{{\varepsilon}}

 \def\slug{\rule{0.1in}{0.1in}}

\begin{document}

\title{\bf Qubit Complexity 
of Continuous Problems\footnote{This research was supported in part 
by DARPA and NSF.}}
\author{A. Papageorgiou}
\email{ap@cs.columbia.edu}
\author{J. F. Traub}
\email{traub@cs.columbia.edu}
\affiliation{Department of Computer Science, Columbia University, 
New York, USA}

\date{\today}

\begin{abstract} 
The number of qubits used by a quantum algorithm will be a crucial
computational resource for the foreseeable future.  We show how to
obtain the classical query complexity for continuous problems. 
We then establish a simple formula for a lower bound
on the qubit complexity in terms of the classical query complexity.
\end{abstract}

\maketitle

\section{Introduction}

There are two major motivations for studying algorithms and complexity of 
continuous problems.
\begin{enumerate} 
\item Many scientific problems have continuous formulations. 
Examples include path integration, Feynman-Kac path integration, and the
Schr\"{o}dinger equation.
\item Are quantum computers more powerful than classical computers for 
important scientific problems? How much more powerful?
\end{enumerate}

To answer these questions one must know the classical computational complexity
of the problem. There are especially constructed problems such as Simon's
problem \cite{S97} 
for which the quantum speedup is known; see also \cite{bennet}. 
Furthermore, it is known that quantum
computers enjoy quadratic speedup for search of an unordered database 
\cite{grover}. Knowing the quantum speedup for such a discrete problem is the 
exception. 
Generally, for discrete problems we do not know the computational
complexity. (Examples of discrete problems are 3-SAT and the traveling salesman
problem.) We have to settle for the conjecture that the complexity hierarchy
does not collapse. A famous example of this conjecture is that $\rm{P}\ne
\rm{NP}$. Thus although is is widely believed
that Shor's algorithm \cite{shor} gives an
exponential speedup it is only a conjecture because the classical 
computational complexity of integer factorization is an important open problem.

In what follows it is important to stress the difference between the cost of 
an algorithm for solving a given problem, 
and the computational complexity of this problem. The
computational complexity (for brevity, the complexity) is the {\it minimal}
computational resources needed to solve the problem. Examples of computational
resources, which have been studied, include memory, time, and communication
on a classical computer and qubits, quantum gates and queries on a 
quantum computer. For the foreseeable future qubits will be a limiting 
resource and in this paper we'll give a general lower bound on the 
qubit complexity for continuous problems.

For continuous problems we often know the classical complexity. There is a 
large literature in the field of
information-based complexity which studies problems with 
partial and/or contaminated information; see \cite{traub,TW98} and the 
references therein. Since functions of a continuous variable cannot generally
be input into a digital computer, the computer has only partial information 
about them. As we shall see in Section II this makes it 
possible to use an adversary argument to get a lower bound on the
classical query complexity, 
and, therefore, on the total complexity of many 
continuous problems.

Most continuous problems arising in practice cannot be solved analytically; 
they must be solved numerically. Since a digital computer has only partial 
information about the input function the problem can only be solved 
approximately, to within an error threshold $\e$. If one insists on an 
error at most $\e$ for all inputs in a class $F$ (the worst case setting) it's 
been shown that for many multivariate problems the complexity is exponential 
in the number of variables. This is known as the curse of dimensionality and
such problems are said to be intractable. Note that for continuous problems
many problems are known to be intractable while for discrete problems
the intractability of NP-hard problems is only conjectured; see 
Remark~\ref{rk:rk24}.

There are two major ways to break the curse of dimensionality, see 
\cite[p.~24]{TW98}. We can weaken the worst case assurance, 
accepting instead a stochastic assurance such as in 
the randomized setting. The Monte Carlo algorithm
is known to be optimal for integration in this setting 
if $F$ is the class of bounded 
continuous functions. Or we can change the class $F$ of inputs. By suitable 
choices of $F$ we can sometimes provide a worst case guarantee while
breaking intractability.

We outline the remainder of the paper. In Section II 
we illustrate the adversary
argument which will provide us with the classical information complexity.
We use a very simple example to do this. In Section III we provide a more
general formulation and introduce notation. In the concluding section we'll 
prove a general theorem giving a lower bound on the qubit complexity in terms 
of the classical query complexity.

\section{Classical information complexity}

We will illustrate the adversary argument used to obtain
the classical query complexity using a very simple example. The same idea can
be applied very generally \cite{traub,TW98}.

\begin{figure}[!h]
\centering
\includegraphics[width=3in, height=0.75in]{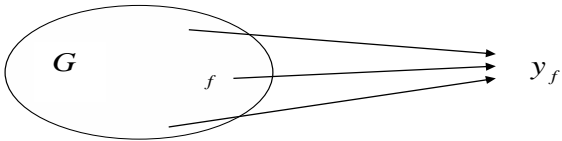}
\caption{Indistinguishable functions}
\label{fig:f1}
\end{figure}

We want to compute $I(f)=\int_0^1 f(x)\,dx$. 
Call $f$ the mathematical input. 
For most integrands we can't
use the fundamental theorem of calculus to compute the integral analytically;
we have to approximate it numerically. 
Although we can input the symbolic form of $f$ into a 
digital computer it doesn't help us to compute the integral. 
We compute 
$$y_f=[f(t_1),\dots, f(t_n)]$$
at $n$ a priori chosen deterministic points $t_i$, $i=1,\dots,n$. Given $y_f$,
there are an infinite number of functions with the same $y_f$. That is, 
we have only partial information about the mathematical input.
Even though the functions may have the same $y_f$ their integrals
may be very different. Let $G$ be the
set of functions with the same $y_f$; see Figure~\ref{fig:f1}.

\begin{figure}[!h]
\centering
\includegraphics[width=3in, height=1.125in]{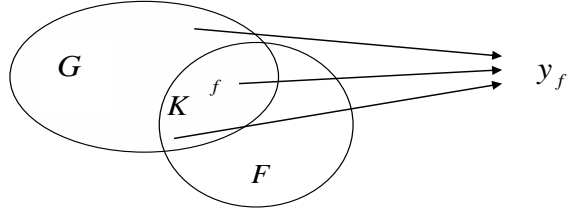}
\caption{Restrict the class of functions}
\label{fig:f2}
\end{figure}

If we only assume that $f$ is, say, Riemann-integrable the classical query
complexity is infinite, i.e, we cannot achieve any desired accuracy no 
matter how large $n$ is.
To get finite complexity we have to make a promise about $f$. 
With the promise that the absolute value of the functions under consideration 
is uniformly bounded by a known constant is 
it is easy to show the complexity is still infinite.
Thus we further restrict the class of inputs and assume that our 
function belongs to
$$F=\left\{ f:\; |f^\prime(x)|\le L, \quad x\in [0,1]\right\}.$$
Let $K=F\cap G$. The functions in $K$ are indistinguishable; 
see Figure \ref{fig:f2}.

Let $H$ denote the
set $I(\tilde f)$, $\tilde f\in K$. It is easy to show that $H$ is an interval
and that its length varies with $y_f$, $n$ and the points $t_1,\dots,t_n$.
Any number in $H$ is a potential approximation to the integral.
A measure of the intrinsic uncertainty in our approximations is the size of 
$H$. There is a standard concept of the size of a set; it is the radius of the
smallest ball containing the set. 
We call this radius the radius of information, $rad$, because its magnitude
depends on how much information we have about the true $f$. It is easy to 
show that we can guarantee an $\e$-approximation iff $rad\le \e$. Let
$m(\e)$ be the minimum number of function evaluations needed 
to solve the problem to within $\e$. The condition $rad\le \e$ implies that
if we compute less than $m(\e)$ function evaluations there does not exist any 
algorithm which solves the problem with error $\e$.
See \cite[Section II.2]{TW98} for a general
discussion of the radius of information.

Let $\mathbf{c}$ be the cost of a query, that is of a function evaluation. 
We define the classical 
query complexity,
$\comp^{\rm query}_{\rm clas}(\e)$, as 
\begin{equation}
\comp^{\rm query}_{\rm clas}(\e)=\mathbf{c}\; m(\e).
\label{eq:i-compl}
\end{equation}
The query complexity is the minimum amount that must be paid to obtain
the information about $f$ needed to compute $I(f)$ to within $\e$.

In the concluding section we will see how the classical query complexity
is used to lower bound the quantum qubit complexity. We conclude this section 
with some remarks.

\begin{rk} The type of argument we have used in this section is called an 
adversary argument because if we don't collect enough information an imagined
adversary can claim the mathematical input is a function $g$ for which
$I(g)$ is very different than $I(f)$, foiling the assurance that we've 
computed an $\e$-approximation to $I(f)$.
\end{rk}

\begin{rk} Note that there has been no mention of how $y_f=[f(t_1),\dots, 
f(t_n)]$ is used to approximate the integral $I(f)$. This can be done by
an algorithm $\phi$ of the form
$$\phi(f)=\sum_{j=1}^na_jf(t_j).$$
There is a large literature on the optimal choice of the coefficients $a_j$
in $\phi$, and on the
optimal choice of the points $t_j$; see, for example
\cite{TW98}. Part of the power of the approach we've illustrated here is that 
decisions concerning information can be separated from 
decisions regarding algorithms.
\end{rk}

\begin{rk} The mathematical tools for lower and upper bounds on classical 
query 
complexity (and other types of complexity) are often deep but this example 
gives the idea of the adversary argument.
\end{rk}

\begin{rk} \label{rk:rk24}
Why can we obtain the complexity of continuous problems
whereas we have to settle for conjectures about the complexity hierarchy 
for discrete problems? 
For continuous problems we have partial information 
and we can use the adversary argument to get lower bounds.
For discrete problems we have complete information. 
For example, for the traveling salesman problem we are given the locations 
of the cities and these coordinates can be input into a digital computer.
There is no information level and no adversary argument.
\end{rk}

\section{Fundamental concepts and notation for quantum computation}

A quantum algorithm consists of a sequence of unitary 
transformations applied to an initial state. The result of
the algorithm is obtained by measuring its final state.
The quantum model of computation is discussed in detail in
\cite{beals,bernstein,bennet,cleve,heinrich,nielsen}. 
We summarize this model to the extent necessary for this paper.

The initial state $|\psi_0\rangle$ of the algorithm is a unit vector 
of the Hilbert space
$\Cal{H}_\nu=\complex^2\otimes \cdots\otimes \complex^2$, $\nu$ times,  
for some appropriately chosen integer $\nu$, where $\complex^2$ is the 
two dimensional space of complex numbers. The dimension of
$\Cal{H}_\nu$ is $2^{\nu}$. The number $\nu$ denotes the number of 
qubits used by the quantum algorithm. 

The final state $|\psi\rangle$ is also a unit vector of
$\Cal{H}_\nu$ and is obtained from the initial state 
$|\psi_0\rangle$ through
a sequence of unitary $2^{\nu}\times 2^{\nu}$ matrices, i.e.,
\begin{equation}
|\psi\rangle\,:=\,U_TQ_fU_{T-1}Q_f\cdots U_1Q_fU_0 |\psi_0\rangle.
\label{eq:qa}
\end{equation}
The unitary matrix $Q_f$  
is called a quantum query and is used to provide information to the
algorithm about a function $f$. 
$Q_f$ depends on $n$ function evaluations $f(t_1),\dots,f(t_n)$,
$n\le 2^{\nu}$. 
The $U_0,U_1,\dots,U_T$ are unitary matrices that do not depend 
on $f$. 
The integer $T$ denotes the number of quantum
queries.

For algorithms solving discrete problems, 
such as Grover's algorithm for the search of an unordered database
\cite{grover}, the input $f$ is considered to be a Boolean function.
However, classical algorithms solving continuous problems using 
floating or fixed point
arithmetic can also be written in the form of 
(\ref{eq:qa}). Indeed, all classical bit operations 
can be simulated by quantum computations, see e.g., \cite{bernstein}. 

The most commonly studied quantum query is the {\it bit} query.
For a Boolean function $f:\{0,\dots,2^m-1\}\to\{0,1\}$, 
the bit query is defined by  
$$
Q_f|j\rangle|k\rangle\,=\,|j\rangle|k\oplus f(j)\rangle. 
$$
Here $\nu=m+1$, $|j\rangle\in\Cal{H}_m$, and
$|k\rangle\in\Cal{H}_{1}$ with $\oplus$ denoting the addition
modulo $2$. For a real function $f$ the query
is constructed by taking the 
most significant bits of the function $f$ evaluated at some points
$t_j$.
More precisely, as in \cite{heinrich}, the bit query for $f$ has the
form
$$
Q_f|j\rangle|k\rangle\,=\,|j\rangle|k\oplus \beta(f(\tau(j)))\rangle, 
$$
where the number of qubits is now $\nu=m'+m''$ and $|j\rangle\in
\Cal{H}_{m'}$, $|k\rangle\in\Cal{H}_{m''}$. The functions
$\beta$ and $\tau$ are used to discretize the domain $\mathcal{D}$ 
and the range $\mathcal{R}$ of $f$, respectively.
Therefore,
$\beta:\mathcal{R}\to\{0,1,\dots,2^{m''}-1\}$ and
$\tau:\{0,1,\dots,2^{m'}-1\}\to\mathcal{D}$.
Hence, we compute $f$ at 
$t_j=\tau(j)$ and then take the $m''$ most significant bits of
$f(t_j)$ by $\beta(f(t_j))$, for the details and the possible use
of ancillary qubits see  \cite{heinrich}.

At the end of the quantum algorithm, a measurement is 
applied to its final state $|\psi \rangle$.
The measurement produces one of $M$ outcomes, where $M\le 2^{\nu}$. 
Outcome $j\in\{0,1,\dots,M-1\}$ occurs with
probability $p_f(j)$, which depends on $j$ and the input $f$.
Knowing the outcome $j$, we compute classically 
the final result $\phi_f(j)$ of the algorithm. 

In principle, quantum algorithms may have many measurements
applied between sequences of unitary transformations of the form 
presented above. However, any algorithm with many measurements 
can be simulated by
a quantum algorithm with only one measurement at the end
\cite{bernstein}.

We are interested in continuous problems such as multivariate and 
path integration,
multivariate approximation, ordinary and partial differential equations, 
and the 
Sturm-Liouville eigenvalue problem. 
For many continuous problems we know tight quantum complexity bounds
\cite{heinrich,H03,H04a,H04b,Kacewicz,N01,PW05,TW02}.  

Let $S$ be a linear or nonlinear operator such that
\begin{equation}
S:\cal{F}\to \cal{G}.
\label{eq:contprob}
\end{equation}
Typically, $\cal{F}$ is a linear space of continuous real functions 
of several variables,
and $\cal{G}$ is a normed linear space.
We wish to approximate $S(f)$ to within $\e$ for $f\in\cal{F}$.
We approximate $S(f)$ using
$n$ function evaluations $f(t_1),\dots,f(t_n)$ at deterministically
and a priori
chosen sample points. The quantum query $Q_f$ encodes this information, 
and the quantum algorithm obtains this information from $Q_f$.

Without
loss of generality, we consider algorithms that approximate $S(f)$ with
probability $p\ge\tfrac34$. The
local error of the quantum algorithm (\ref{eq:qa})
that computes the approximation $\phi_f(j)$, for
$f\in \cal{F}$ and the outcome $j\in\{0,1,\dots,M-1\}$, is defined by 
\begin{equation}
e(\phi_f)\,=\,\min \bigg\{\, \a :\quad \sum_{j:\ \|S(f) - 
\phi_f(j)\| \,\le\, \a\,}p_f(j)\geq \tfrac34\,\bigg\},
\label{eq:localperr}
\end{equation}
where $p_f(j)$ denotes the probability of obtaining outcome $j$ 
for the function $f$.
The {\it worst probabilistic} error of a quantum algorithm $\phi$ 
is defined by
\begin{equation}
e^{\w}(\phi)\,=\,\sup_{f\in \cal{F}}
e(\phi_f).
\label{eq:wperr}
\end{equation}

\section{Lower bound on qubit complexity}

For the foreseeable future
the number of qubits used by a quantum algorithm will be 
a crucial computational
resource. We will show how to obtain a lower bound for the number of 
qubits needed for algorithms that approximate continuous problems such as
(\ref{eq:contprob}). 
In particular, let 
$\comp^{\rm qubit}(\e)$ be the minimal number of qubits required by a quantum
algorithm of the form (\ref{eq:qa}) approximating $S(f)$ with accuracy $\e$
and probability at least $\tfrac 34$. 

We will derive a lower bound for the qubit complexity 
using facts about the classical complexity of continuous 
problems. 
A similar lower bound result was announced by H. Wo\'zniakowski at the DARPA PI
meeting in Chicago in May 2004; see \cite{W05} for his proof. 
The proof we present
here is different and constructive.
In the analysis of classical algorithms one considers
the classical query cost, which depends on the 
number of function evaluations $n$ used by
the classical algorithm. 
It suffices to consider deterministic classical algorithms $\phi$ 
in the worst case, i.e., to measure the error by
\begin{equation}
e^{\rm wor}(\phi,n)=\sup_{f\in F} \big\| S(f)- \phi(f(t_1),\dots,f(t_n)) 
\big\|.
\label{eq:wcerr}
\end{equation}
The classical query complexity, 
$\comp^{\rm query}_{\rm clas}(\e)$, of the problem 
(\ref{eq:contprob})
is the minimal number of function evaluations that are necessary for
accuracy $\e$ times the cost of a query, i.e.,
\begin{equation*}
m(\e) = \min\left\{ n: \; \exists\; \phi {\rm\ with\ } 
e^{\rm wor}(\phi,n) \le\e \right\}, \\
\end{equation*}
\begin{equation}
\comp^{\rm query}_{\rm clas}(\e) = \mathbf{c}\; m(\e).
\label{eq:infocompl}
\end{equation}
The classical query and combinatorial complexities of many continuous 
problems are known \cite{traub,TW98}. We are now ready to show how to
use classical query complexity lower bounds to derive qubit 
complexity lower bounds.

Recall that
quantum algorithms may require some 
classical computations to be performed, for instance, at the end
after the measurement to produce the final result,  
or at the beginning to prepare the initial state.
These classical computations may or may not 
include a number of function evaluations.
To exclude trivial cases that reduce
the qubit complexity at the expense of classical computations, we will assume
that the number of function evaluations computed by the classical components
of the quantum algorithm cannot exceed the number of function evaluations
obtained in superposition by the query due to quantum parallelism.

\begin{thm} The qubit complexity of a quantum algorithm (\ref{eq:qa}) that
solves the problem (\ref{eq:contprob}) with accuracy $\e$ is bounded from
below as follows
\begin{equation*}
\comp^{\rm qubit}(\e) \ge \log_2 \left[ \comp^{\rm query}_{\rm clas}
(3\e)\right]- 1.
\end{equation*}
\end{thm}

\noindent{\bf Proof:} 
Consider a quantum algorithm that solves the problem with accuracy $\e$.
This algorithm uses $Q_f$ which, in turn, 
depends on a number of function 
evaluations of $f$ which we denote by $n(\e)$. It follows that the number of 
qubits of the quantum algorithm is at least $\log_2 n(\e)$.

A quantum algorithm 
that approximates (\ref{eq:contprob}) with accuracy $\e$ 
can be simulated by a classical algorithm. 
The computational cost of this simulation is not important here.
The important fact is that
the classical algorithm also uses $n(\e)$ function evaluations
and approximates $S(f)$ with worst probabilistic error 
(\ref{eq:wperr}) less than $\e$. 

Since the algorithm achieves error $\e$, 
the final state of the quantum algorithm, and the corresponding
state of its classical simulation,
contain outcomes $j$ such that $\| S(f)-\phi_f(j)\|\le \e$, where
the sum of their probabilities is 
$\sum_j p_f(j)\ge \tfrac 34$. Moreover, the classical simulation 
can compute the probabilities of all the outcomes,
since it has computed all the amplitudes in the final state of 
the quantum algorithm.

The quantities $p_f(j)$ and $\phi_f(j)$, for all possible outcomes $j$,
suffice for computing deterministically an approximation of $S(f)$ with
error $3\e$. To see this observe that the local error 
(\ref{eq:localperr}) of a quantum algorithm
can be equivalently rewritten as
\begin{equation}
e(\phi_f)\,=\,\min_{A:\, \mu(A)\ge \tfrac34}\max_{j\in A}
\big\| S(f)-\phi_f(j)\big\|,
\label{eq:localperr2}
\end{equation}
where $A\subset\{0,1,\dots,M-1\}$ and $\mu(A)=\sum_{j\in A}p_f(j)$. 
Consider all sets of outcomes where the sum of the respective probabilities
is at least $\tfrac 34$. From these discard any set that contains 
outcomes $j\ne k$
such that $\|\phi_f(j)-\phi_f(k)\| > 2\e$. 

Let $A$ denote one of the remaining
sets of outcomes then $\mu(A^*)=\sum_{j\in A^*}p_f(j) \ge\tfrac 34$ and
$\|\phi_f(j)-\phi_f(k)\| \le 2\e$, $j,k\in A$. The fact that $e(\phi_f)\le\e$,
equation (\ref{eq:localperr2}) 
and the triangle inequality imply that $A$ exists.
 
There exists $j^*\in A$ such that
$\|S(f)- \phi_f(j^*)\| \le \e$. Indeed, if we assume that
$\|S(f)- \phi_f(j)\| > \e$,
for all $j\in A$, then the quantum algorithm cannot have accuracy $\e$ 
with probability 
at least $\tfrac 34$, and we reach a contradiction. 

The triangle 
inequality yields that $\|S(f)- \phi_f(j)\|
\le \|S(f)-\phi_f(j^*)\| + \|\phi_f(j^*)-\phi_f(j)\| \le 3\e$, 
for any $j\in A$.
Hence, we have obtained a deterministic classical algorithm that solves
the problem with error $3\e$.

By our assumption, the classical components of the quantum 
algorithm may contain a number of function evaluations up to $n(\e)$
which implies
\begin{equation}
2n(\e)\ge \comp^{\rm query}_{\rm clas}(3\e).
\label{eq:lb2}
\end{equation}

Since the quantum algorithm must have at least $\log_2 n(\e)$ qubits,
as we indicated at the beginning of the proof, equation (\ref{eq:lb2})
implies that the qubit complexity of the quantum algorithm is bounded from 
below as follows
\begin{equation*}
\comp^{\rm qubit}(\e) \ge \log_2 n(\e) \ge
\log_2 \left[ \comp^{\rm query}_{\rm clas}(3\e)\right]-1
\end{equation*}
and the proof is complete.\hfill\slug

\vskip 1pc

As we have already indicated quantum algorithms may have several measurements.
They are sequences of quantum algorithms with a single measurement,
i.e., a sequences of algorithms of the form (\ref{eq:qa}), and the resulting 
algorithm has success probability, say, $\tfrac 34$. 
The individual quantum algorithms may use different numbers of qubits,
and we denote by $k$ the maximum of these numbers. One may reduce
$k$ not only at the expense of classical function evaluations but also
by considering extremely long sequences of quantum algorithms with
a single measurement. Therefore, to exclude such trivial cases we will
assume that the total number of classical function evaluations used
by the classical components of a sequence of quantum algorithms 
is a polynomial in $2^k$, and so is the number of quantum algorithms
with a single measurement that have been combined together to form
the quantum algorithm with several measurements. Under these conditions 
we have the following corollary.

\begin{cor}The qubit complexity of a quantum algorithm with several
measurements is bounded as 
$$
\comp^{\rm qubit}(\e) =\Omega\left(  
\log_2 \left[ \comp^{\rm query}_{\rm clas}(3\e)\right]\right).
$$ 
\end{cor}

\bibliography{qbcomp}

\begin{thebibliography}{19}
\expandafter\ifx\csname natexlab\endcsname\relax\def\natexlab#1{#1}\fi
\expandafter\ifx\csname bibnamefont\endcsname\relax
  \def\bibnamefont#1{#1}\fi
\expandafter\ifx\csname bibfnamefont\endcsname\relax
  \def\bibfnamefont#1{#1}\fi
\expandafter\ifx\csname citenamefont\endcsname\relax
  \def\citenamefont#1{#1}\fi
\expandafter\ifx\csname url\endcsname\relax
  \def\url#1{\texttt{#1}}\fi
\expandafter\ifx\csname urlprefix\endcsname\relax\def\urlprefix{URL }\fi
\providecommand{\bibinfo}[2]{#2}
\providecommand{\eprint}[2][]{\url{#2}}

\bibitem[{\citenamefont{Simon}(1997)}]{S97}
\bibinfo{author}{\bibfnamefont{D.~R.} \bibnamefont{Simon}},
  \bibinfo{journal}{SIAM J. Comput.} \textbf{\bibinfo{volume}{26}},
  \bibinfo{pages}{1474} (\bibinfo{year}{1997}).

\bibitem[{\citenamefont{Bennett et~al.}(1997)\citenamefont{Bennett, Bernstein,
  Brassard, and Vazirani}}]{bennet}
\bibinfo{author}{\bibfnamefont{C.~H.} \bibnamefont{Bennett}},
  \bibinfo{author}{\bibfnamefont{E.}~\bibnamefont{Bernstein}},
  \bibinfo{author}{\bibfnamefont{G.}~\bibnamefont{Brassard}}, \bibnamefont{and}
  \bibinfo{author}{\bibfnamefont{U.}~\bibnamefont{Vazirani}},
  \bibinfo{journal}{SIAM J. Computing} \textbf{\bibinfo{volume}{26(5)}},
  \bibinfo{pages}{1510} (\bibinfo{year}{1997}).

\bibitem[{\citenamefont{Grover}(1997)}]{grover}
\bibinfo{author}{\bibfnamefont{L.}~\bibnamefont{Grover}},
  \bibinfo{journal}{Phys. Rev. Lett.} \textbf{\bibinfo{volume}{79(2)}},
  \bibinfo{pages}{325} (\bibinfo{year}{1997}), \eprint{quant-ph/9706033}.

\bibitem[{\citenamefont{Shor}(1997)}]{shor}
\bibinfo{author}{\bibfnamefont{P.~W.} \bibnamefont{Shor}},
  \bibinfo{journal}{SIAM J. Comput.} \textbf{\bibinfo{volume}{26(5)}},
  \bibinfo{pages}{1484} (\bibinfo{year}{1997}).

\bibitem[{\citenamefont{Traub et~al.}(1988)\citenamefont{Traub, Wasilkowski,
  and Wo\'zniakowski}}]{traub}
\bibinfo{author}{\bibfnamefont{J.~F.} \bibnamefont{Traub}},
  \bibinfo{author}{\bibfnamefont{G.~W.} \bibnamefont{Wasilkowski}},
  \bibnamefont{and}
  \bibinfo{author}{\bibfnamefont{H.}~\bibnamefont{Wo\'zniakowski}},
  \emph{\bibinfo{title}{Information-Based Complexity}}
  (\bibinfo{publisher}{Academic Press}, \bibinfo{year}{1988}).

\bibitem[{\citenamefont{Traub and Werschulz}(1998)}]{TW98}
\bibinfo{author}{\bibfnamefont{J.~F.} \bibnamefont{Traub}} \bibnamefont{and}
  \bibinfo{author}{\bibfnamefont{A.~G.} \bibnamefont{Werschulz}},
  \emph{\bibinfo{title}{Complexity and Information}}
  (\bibinfo{publisher}{Cambridge University Press}, \bibinfo{year}{1998}).

\bibitem[{\citenamefont{Beals et~al.}(1998)\citenamefont{Beals, Buhrman, Cleve,
  Mosca, and de~Wolf}}]{beals}
\bibinfo{author}{\bibfnamefont{R.}~\bibnamefont{Beals}},
  \bibinfo{author}{\bibfnamefont{H.}~\bibnamefont{Buhrman}},
  \bibinfo{author}{\bibfnamefont{R.}~\bibnamefont{Cleve}},
  \bibinfo{author}{\bibfnamefont{M.}~\bibnamefont{Mosca}}, \bibnamefont{and}
  \bibinfo{author}{\bibfnamefont{R.}~\bibnamefont{de~Wolf}},
  \bibinfo{journal}{Proceedings FOCS'98} p. \bibinfo{pages}{352}
  (\bibinfo{year}{1998}), \eprint{quant-ph/9802049}.

\bibitem[{\citenamefont{Bernstein and Vazirani}(1997)}]{bernstein}
\bibinfo{author}{\bibfnamefont{E.}~\bibnamefont{Bernstein}} \bibnamefont{and}
  \bibinfo{author}{\bibfnamefont{U.}~\bibnamefont{Vazirani}},
  \bibinfo{journal}{SIAM J. Computing} \textbf{\bibinfo{volume}{26(5)}},
  \bibinfo{pages}{1411} (\bibinfo{year}{1997}).

\bibitem[{\citenamefont{Cleve et~al.}(1996)\citenamefont{Cleve, Ekert,
  Macchiavello, and Mosca}}]{cleve}
\bibinfo{author}{\bibfnamefont{R.}~\bibnamefont{Cleve}},
  \bibinfo{author}{\bibfnamefont{A.}~\bibnamefont{Ekert}},
  \bibinfo{author}{\bibfnamefont{C.}~\bibnamefont{Macchiavello}},
  \bibnamefont{and} \bibinfo{author}{\bibfnamefont{M.}~\bibnamefont{Mosca}},
  \bibinfo{journal}{Phil. Trans. R. Soc. Lond. A.}  (\bibinfo{year}{1996}).

\bibitem[{\citenamefont{Heinrich}(2002)}]{heinrich}
\bibinfo{author}{\bibfnamefont{S.}~\bibnamefont{Heinrich}},
  \bibinfo{journal}{J. Complexity} \textbf{\bibinfo{volume}{18(1)}},
  \bibinfo{pages}{1} (\bibinfo{year}{2002}), \eprint{quant-ph/0105116}.

\bibitem[{\citenamefont{Nielsen and Chuang}(2000)}]{nielsen}
\bibinfo{author}{\bibfnamefont{M.~A.} \bibnamefont{Nielsen}} \bibnamefont{and}
  \bibinfo{author}{\bibfnamefont{I.~L.} \bibnamefont{Chuang}},
  \emph{\bibinfo{title}{Quantum Computation and Quantum Information}}
  (\bibinfo{publisher}{Cambridge University Press}, \bibinfo{year}{2000}).

\bibitem[{\citenamefont{Heinrich}(2003)}]{H03}
\bibinfo{author}{\bibfnamefont{S.}~\bibnamefont{Heinrich}},
  \bibinfo{journal}{J. Complexity} \textbf{\bibinfo{volume}{19}},
  \bibinfo{pages}{19} (\bibinfo{year}{2003}).

\bibitem[{\citenamefont{Heinrich}(2004{\natexlab{a}})}]{H04a}
\bibinfo{author}{\bibfnamefont{S.}~\bibnamefont{Heinrich}},
  \bibinfo{journal}{J. Complexity} \textbf{\bibinfo{volume}{20}},
  \bibinfo{pages}{5} (\bibinfo{year}{2004}{\natexlab{a}}),
  \eprint{quant-ph/0305030}.

\bibitem[{\citenamefont{Heinrich}(2004{\natexlab{b}})}]{H04b}
\bibinfo{author}{\bibfnamefont{S.}~\bibnamefont{Heinrich}},
  \bibinfo{journal}{J. Complexity} \textbf{\bibinfo{volume}{20}},
  \bibinfo{pages}{27} (\bibinfo{year}{2004}{\natexlab{b}}),
  \eprint{quant-ph/0305031}.

\bibitem[{\citenamefont{Kacewicz}(2004)}]{Kacewicz}
\bibinfo{author}{\bibfnamefont{B.~Z.} \bibnamefont{Kacewicz}},
  \bibinfo{journal}{J. Complexity} \textbf{\bibinfo{volume}{21(5)}},
  \bibinfo{pages}{740} (\bibinfo{year}{2004}).

\bibitem[{\citenamefont{Novak}(2001)}]{N01}
\bibinfo{author}{\bibfnamefont{E.}~\bibnamefont{Novak}}, \bibinfo{journal}{J.
  Complexity} \textbf{\bibinfo{volume}{17}}, \bibinfo{pages}{2}
  (\bibinfo{year}{2001}), \eprint{quant-ph/0008124}.

\bibitem[{\citenamefont{Papageorgiou and Wo\'zniakowski}(2005)}]{PW05}
\bibinfo{author}{\bibfnamefont{A.}~\bibnamefont{Papageorgiou}}
  \bibnamefont{and}
  \bibinfo{author}{\bibfnamefont{H.}~\bibnamefont{Wo\'zniakowski}},
  \bibinfo{journal}{Quantum Information Processing}
  \textbf{\bibinfo{volume}{4(2)}}, \bibinfo{pages}{87} (\bibinfo{year}{2005}),
  \eprint{quant-ph/0502054}.

\bibitem[{\citenamefont{Traub and Wo\'zniakowski}(2002)}]{TW02}
\bibinfo{author}{\bibfnamefont{J.~F.} \bibnamefont{Traub}} \bibnamefont{and}
  \bibinfo{author}{\bibfnamefont{H.}~\bibnamefont{Wo\'zniakowski}},
  \bibinfo{journal}{Quantum Information Processing}
  \textbf{\bibinfo{volume}{1(5)}}, \bibinfo{pages}{365} (\bibinfo{year}{2002}),
  \eprint{quant-ph/0109113}.

\bibitem[{\citenamefont{Wo\'zniakowski}(2005)}]{W05}
\bibinfo{author}{\bibfnamefont{H.}~\bibnamefont{Wo\'zniakowski}},
  \emph{\bibinfo{title}{The quantum setting with randomized queries for
  continuous problems}} (\bibinfo{year}{2005}), \bibinfo{note}{in progress}.

\end{thebibliography}

\end{document}